\documentclass[12pt]{article}

\newcommand{\eq}[1]{Eq.~(\ref{#1})}

\def\beq{\begin{equation}}
\def\eeq{\end{equation}}
\def\beqa{\begin{eqnarray}}
\def\eeqa{\end{eqnarray}}

\newcommand{\EQ}{\begin{equation}}
\newcommand{\EN}{\end{equation}}
\newcommand{\bea}{\begin{eqnarray}}
\newcommand{\ena}{\end{eqnarray}}

\renewcommand{\a}{\alpha}
\renewcommand{\b}{\beta}

\newcommand{\z}{\zeta}

\newcommand{\NP}[1]{Nucl.\ Phys.\ {\bf #1}}
\newcommand{\PL}[1]{Phys.\ Lett.\ {\bf #1}}

\def\one{{\hbox{ 1\kern-.8mm l}}}

\def\tr{{\rm tr\,}}

\newlength{\bredde}
\def\slash#1{\settowidth{\bredde}{$#1$}\ifmmode\,\raisebox{.15ex}{/}
\hspace*{-\bredde} #1\else$\,\raisebox{.15ex}{/}\hspace*{-\bredde} #1$\fi}

\begin{document}\begin{titlepage}
\begin{flushright} KUL-TF-98/58 \\ hep-th/9812149
\end{flushright}
\vfill
\begin{center} 
{\LARGE\bf (Non-)Anomalous D-brane and O-plane couplings: the normal bundle}    \\
\vskip 27.mm  \large
{\bf   Ben Craps $^{1,2}$, Frederik Roose $^3$ } \\
\vskip 1cm
{\em Instituut voor theoretische fysica}\\
{\em Katholieke Universiteit Leuven, B-3001 Leuven, Belgium}
\end{center}
\vfill

\begin{center}
{\bf ABSTRACT}
\end{center}
\begin{quote}
The direct string computation of anomalous D-brane and orientifold plane
couplings is extended to include the curvature of the normal bundle. 
The normalization of these terms is fixed unambiguously. New, non-anomalous
gravitational couplings are found.

\vfill      \hrule width 5.cm
\vskip 2.mm
{\small
\noindent $^1$ Aspirant FWO, Belgium }\\
{\small
\noindent $^2$ E-mail: Ben.Craps@fys.kuleuven.ac.be }\\
{\small
\noindent $^3$ E-mail: Frederik.Roose@fys.kuleuven.ac.be }
\end{quote}
\begin{flushleft}
\end{flushleft}
\end{titlepage}

\section{Introduction}
The Wess-Zumino part of the effective D-brane action has the following form,
which can be derived from anomaly cancellation arguments
\cite{GHM, CY}:
\beq
\label{WZaction}
S_{WZ}=\frac{T_p}{\kappa}\int_{p+1}\hat{{\cal C}'}\wedge e^{2\pi\a '\,F+\hat{B'}}\wedge
\sqrt{\hat{A}(R_T)/\hat{A}(R_N)}~~.
\eeq
Here $R_T$ and $R_N$ are the curvatures of the tangent and normal bundles of
the D-brane worldvolume, and $\hat{A}$ denotes the A-roof genus:
\beqa
\label{Aroof}
\sqrt{\frac{\hat{A}(R_T)}{\hat{A}(R_N)}}&=&1+\frac{(4\pi^2\a ')^2}{384\pi^2}(\tr R_T^2-\tr R_N^2)
+\frac{(4\pi^2\a ')^4}{294912\pi^4}(\tr R_T^2-\tr R_N^2)^2\nonumber\\&&+
\frac{(4\pi^2\a ')^4}{184320\pi^4} (\tr R_T^4-\tr R_N^4)+\ldots
\eeqa
For the other symbols and our conventions we refer to 
Ref.~\cite{benfred}. Here we will be interested in the gravitational
part of these interactions, in particular in the part involving the normal
bundle.

Similar interactions have been proposed for orientifold planes, which do not
have a gauge field on their worldvolumes \cite{Das, MSS}:
\beq
\label{Oaction}
S_{WZ}=-2^{p-4}\frac{T_p}{\kappa}\int_{p+1}\hat{{\cal C}'}\wedge
\sqrt{L(R_T/4)/L(R_N/4)}~~.
\eeq
The Hirzebruch polynomials $L$ are given by:
\beqa
\label{hirzebruch}
\sqrt{\frac{L(R_T/4)}{L(R_N/4)}}&=&1-\frac{(4\pi^2\a ')^2}{768\pi^2}(\tr R_T^2-\tr R_N^2)
+\frac{(4\pi^2\a ')^4}{1179648\pi^4}(\tr R_T^2-\tr R_N^2)^2\nonumber\\&&-
\frac{7(4\pi^2\a ')^4}{1474560\pi^4} (\tr R_T^4-\tr R_N^4)+\ldots
\eeqa

Until a few months ago, all the evidence in favour of the gravitational terms
in \eq{WZaction} and \eq{Oaction} consisted of duality \cite{BerVaf} and 
anomaly arguments. Recently there have been various more explicit checks.

In Ref.~\cite{benfred} the presence of the four-form  gravitational couplings 
involving the tangent bundle was verified by explicit disc and crosscap 
computations in string theory.

All the couplings in \eq{WZaction} and \eq{Oaction} were computed in 
Ref.~\cite{MSS} by factorizing various one-loop amplitudes in the RR channel.
In particular, the 8-form gravitational coupling on O-planes was corrected and
the topological origin of these terms was clarified. However, being less 
direct than the disc/crosscap computations, the method that was used
does not allow, for instance, to determine 
conclusively the normalization of the normal
bundle interaction. The reason is that their argument depends on formal
manipulations of vanishing amplitudes.

In Ref.~\cite{stefanski} the direct method of Ref.~\cite{benfred} was extended
to include all the tangent bundle couplings present in \eq{WZaction} and 
\eq{Oaction}. Agreement was found with the relevant results of Ref.~\cite{MSS}.
Moreover it was shown that the integral determining the normalization of the
4-form coupling computed in Ref.~\cite{benfred} is in fact a factor two smaller
than thought before, thus solving the apparent discrepancy between the string
and supergravity calculations in the latter paper.

To summarize, the project of determining the anomalous gravitational D-brane
and O-plane couplings by computing string scattering amplitudes
seems to be almost completed. Only a direct computation
of the normal bundle couplings has been missing. Besides having the 
conceptual advantage of being direct and unrelated to the earlier checks 
(which relied
on anomaly arguments), it would also allow one to fix the normalizations in an
unambiguous way. 

The aim of the present paper is to provide this direct computation of the
normal bundle couplings. In fact, it will be shown that they can be extracted
rather easily from the computations in Refs~\cite{benfred, stefanski}. In those
papers the restriction was made at some point to graviton
polarizations and momenta purely
along the brane/plane. This simplified the integrals to be performed. The main
point of the present paper is that this restriction need not be made: analogous
simplifications occur when restricting to small momenta, leaving the
polarizations and momenta otherwise arbitrary. 

The result will be that all anomalous gravitational D-brane and O-plane
couplings are reproduced. In particular, also the normal bundle couplings come
with the expected coefficients.

It turns out that the WZ-terms of brane and plane effective actions also
contain non-anomalous, eightform gravitational terms. As far as we know, these
have not been noted before. 
  
\section{The two graviton amplitude}
In a recent paper \cite{benfred} the amplitude for two gravitons and one RR potential
in the
presence of a D-brane was calculated. With the conventions and normalizations adopted in 
that paper the amplitude was shown to be
\begin{eqnarray}
{\cal A}&=&
\frac{\kappa^2\,T_p\a '^2}{4\sqrt{2}\,(p-3)!\,\pi^2}\,
\epsilon^{\a_1\cdots\a_{p-3}\b_1\cdots\b_4}\,c_{\a_1\cdots\a_{p-3}}\,
k_{3\b_1}\,k_{4\b_3}\nonumber\\ 
&&[(k_4\cdot S\cdot\zeta_{3\b_2})(k_3\cdot\zeta_{4\b_4})
-(k_3\cdot S\cdot k_4)(\zeta_{3\b_2}
\cdot\zeta_{4\b_4})
+(k_4\cdot\zeta_{3\b_2})(k_3\cdot S\cdot\zeta_{4\b_4})\nonumber \\ 
&&-(k_3\cdot k_4)(\zeta_{3\b_2}\cdot S\cdot\zeta_{4\b_4})]\nonumber\\
&&\int_{|z_3|,|z_4|>1}d^2z_3\,d^2z_4\,
(|z_3|^2-1)^{k_3\cdot S\cdot k_3}\,(|z_4|^2-1)^{k_4\cdot S\cdot k_4}\,
|z_3|^{-2\,k_3\cdot S\cdot k_3-2\,k_3\cdot S\cdot k_4-2} \nonumber \\ &&
|z_4|^{-2\,k_4\cdot S\cdot k_4-2\,k_3\cdot S\cdot k_4-2}\, 
|z_3-z_4|^{2\,k_3\cdot k_4-2}\,|z_3\bar z_4-1|^{2\,k_3\cdot S\cdot k_4-2}\,
(z_3\bar z_4-\bar z_3z_4)^2~~.\nonumber
\end{eqnarray}
Restricting to small momenta, the momenta in the exponents can be put to
zero.
As has been shown recently in Ref. \cite{stefanski}, the integral 
then correctly evaluates to 
$\frac{2 \pi^4}{3}$,
which is half the result given in \cite{benfred}. For comparison with supergravity it
will be useful to expand the kinematical prefactor 
in quantities that contain only longitudinal or transverse polarisations 
and momenta. The amplitude is thus found to be
\begin{eqnarray} \label{stramp2}
{\cal A}_{\rm string} &=&
\frac{\sqrt{2}\,\pi^2\,\kappa^2\,T_p\a '^2}{6\,(p-3)!}\,
\epsilon^{\a_1\cdots\a_{p-3}\b_1\cdots\b_4}\,c_{\a_1\cdots\a_{p-3}}\,k_{3\b_1}\,k_{4\b_3} \\ \nonumber
&&[(k^{\Vert}_4\cdot\zeta^{\Vert}_{3\b_2})(k^{\Vert}_3\cdot \zeta^{\Vert}_{4\b_4}) -
 (k^{\Vert}_4\cdot k^{\Vert}_{3})(\zeta^{\Vert}_{3\b_2}\cdot \zeta^{\Vert}_{4\b_4})\\ \nonumber 
&&-(k^{\perp}_4\cdot\zeta^{\perp}_{3\b_2})(k^{\perp}_3\cdot \zeta^{\perp}_{4\b_4})
+(k^{\perp}_4\cdot k^{\perp}_{3})(\zeta^{\perp}_{3\b_2}\cdot \zeta^{\perp}_{4\b_4})]~~,
\end{eqnarray}
that is, there is a split into `tangent' and `normal' pieces relative to the D-brane
worldvolume. 

Let us now check that with the adopted normalizations the string amplitude reproduces the
result as expected from the D-brane action.  
Taking the Wess-Zumino action \eq{WZaction}
the field
theory amplitude evaluates to
\begin{eqnarray}
{\cal A}_{\rm sugra}&=&\frac{\sqrt{2}\,\pi^2\,\kappa^2\,T_p\a '^2}{6\,(p-3)!}\,
\epsilon^{\a_1\cdots\a_{p-3}\b_1\cdots\b_4}\,c_{\a_1\cdots\a_{p-3}}\,k_{3\b_1}\,k_{4\b_3} \\ \nonumber
&&[(k^{\Vert}_4\cdot\zeta^{\Vert}_{3\b_2})(k^{\Vert}_3\cdot \zeta^{\Vert}_{4\b_4}) -
 (k^{\Vert}_4\cdot k^{\Vert}_{3})(\zeta^{\Vert}_{3\b_2}\cdot \zeta^{\Vert}_{4\b_4})\\ \nonumber 
&&-(k^{\perp}_4\cdot\zeta^{\perp}_{3\b_2})(k^{\perp}_3\cdot \zeta^{\perp}_{4\b_4})
+(k^{\perp}_4\cdot k^{\perp}_{3})(\zeta^{\perp}_{3\b_2}\cdot \zeta^{\perp}_{4\b_4})]~~,
\end{eqnarray}
which exactly coincides with the string theory expression \eq{stramp2}. From the string theory
point of view it is rather remarkable how the boundary state encodes the different ways in
which the tangent and normal bundle curvatures appear. 

Let us now combine the above observations and the analysis in 
Ref.~\cite{benfred}
for the orientifold case. Considerations in a recent paper \cite{MSS} suggest 
that the
D-brane Wess-Zumino action  \eq{WZaction} should be replaced by \eq{Oaction} 
for orientifolds.

First, there is a factor $-2^{p-4}$ in front of the orientifold action. This 
accounts for the charge of
orientifold planes in the corresponding D-brane charge units. 
For the O9-plane this factor $-32$ goes into the normalization of the cross-cap 
state. The relative factor
$-\frac 1 2$ between the two-graviton pieces in \eq{Aroof} and \eq{hirzebruch} 
is reflected by the same
proportionality factor between the integral in the first formula of this
section and the corresponding one for the orientifold plane \cite{benfred, stefanski}. 
When considering 
lower-dimensional planes, we
can clearly repeat the above analysis. The normalization of the cross-cap state 
acquires the appropriate
powers of 2, and the kinematical prefactor in the string amplitude nicely 
splits up into tangent and normal pieces.

\section{The four graviton amplitude}
In Ref.~\cite{stefanski} the four graviton amplitude was computed for the
special case of momenta and polarizations along the brane/plane. In this
section we extend his computation to the general case. We will find that the
anomalous couplings (\ref{WZaction}) are correctly reproduced. On top of those 
we find extra gravitational couplings, which cannot be
factorized in terms of the curvatures of the tangent and the normal bundles.

The scattering amplitude of four gravitons and one RR-potential in the
presence of a (high enough) D-brane should reproduce the last two terms of
\eq{Aroof}. The first one of those is the easier: the computation is
essentially the `square' of the one in the previous section. We will not
repeat it here; the result is that this term is precisely reproduced by the part of
the string amplitude for which the gravitons are contracted in pairs, such
that the kinematical factor (and the integral for small momenta!) factorizes
between two pairs of gravitons. (For momenta and polarizations along the
brane, this part of the amplitude corresponds to
the first line of Eq.~(47) in Ref.~\cite{stefanski}.)

The last term of \eq{Aroof} should be reproduced by the non-factorizable
parts of the amplitude (corresponding to the last three lines of
Eq.~(47) in Ref.~\cite{stefanski}). It will turn out that this part of the
amplitude indeed reproduces the anomalous term, but it does more: it also
describes new, non-anomalous gravitational interactions on D-branes.

Without loss of generality, we concentrate on one of the terms (the others
can be obtained by permuting the gravitons and antisymmetrizing in
polarizations and momenta of each graviton). Normalized as in
Ref.~\cite{benfred} it reads
\beqa \label{trR4}
&&\frac{\sqrt{2}T_p \pi^4\kappa^4\a'^4}{128 (p-7)!}\epsilon^{\a_1\ldots\a_{p-7}
\mu_1\ldots\mu_8}c_{\a_1\ldots\a_{p-7}}\z_{1\mu_1}k_{1\mu_2}\ldots\z_{4\mu_7}
k_{4\mu_8}\nonumber\\&&
\times {\bf\{}\frac{14\pi^8}{45}
       [(\z_1\cdot S\cdot k_2)(\z_2\cdot k_3)(\z_3\cdot k_4)(\z_4\cdot k_1)
       +(\z_1\cdot k_2)(\z_2\cdot S\cdot k_3)(\z_3\cdot k_4)(\z_4\cdot k_1)
       \nonumber\\&& \left.
       +(\z_1\cdot k_2)(\z_2\cdot k_3)(\z_3\cdot S\cdot k_4)(\z_4\cdot k_1)
       +(\z_1\cdot k_2)(\z_2\cdot k_3)(\z_3\cdot k_4)(\z_4\cdot S\cdot k_1)]\right.
       \nonumber\\ &&\left.
       +\frac{2\pi^8}{45}
       [(\z_1\cdot S\cdot k_2)(\z_2\cdot S\cdot k_3)(\z_3\cdot S\cdot k_4)(\z_4\cdot k_1)
       \right.\nonumber\\&&\left.
       +(\z_1\cdot S\cdot k_2)(\z_2\cdot S\cdot k_3)(\z_3\cdot k_4)(\z_4\cdot S\cdot k_1)
       \right.\nonumber\\&&\left.
       +(\z_1\cdot S\cdot k_2)(\z_2\cdot k_3)(\z_3\cdot S\cdot k_4)(\z_4\cdot S\cdot k_1)
       \right.\nonumber\\&&
       +(\z_1\cdot k_2)(\z_2\cdot S\cdot k_3)(\z_3\cdot S\cdot k_4)(\z_4\cdot S\cdot k_1)]
       {\bf\}}
\eeqa
(Actually, it can be seen from Eq.~(47) in Ref.~\cite{stefanski} that this should be the
structure: the number of $B_{ij}$'s equals the number of left-right contractions giving
rise to the integral. Compare with Eqs~(2.9) and (2.10) in Ref.~\cite{benfred}.) We refer to
Ref.~\cite{stefanski} for the numerical factors inside the braces, which result from
evaluating a four (complex) dimensional integral analogous to the one appearing at the
beginning of the
previous section.

Writing, as in the previous section,
\beqa
\z_i\cdot k_j&=&\z_i^{\Vert}\cdot k_j^{\Vert}+\z_i^{\perp}\cdot k_j^{\perp}\nonumber\\
\z_i\cdot S\cdot k_j&=&\z_i^{\Vert}\cdot k_j^{\Vert}-\z_i^{\perp}\cdot k_j^{\perp}~~,
\eeqa
it is clear that the expression in braces contains 
\beq
\frac{64\pi^8}{45} (\z_1^{\Vert}\cdot k_2^{\Vert})(\z_2^{\Vert}\cdot k_3^{\Vert})
                   (\z_3^{\Vert}\cdot k_4^{\Vert})(\z_4^{\Vert}\cdot k_1^{\Vert})~,
\eeq
the part derived in Ref.~\cite{stefanski}, and
\beq
-\frac{64\pi^8}{45} (\z_1^{\perp} \cdot k_2^{\perp})(\z_2^{\perp}\cdot k_3^{\perp})
                   (\z_3^{\perp}\cdot k_4^{\perp})(\z_4^{\perp}\cdot k_1^{\perp})~,
\eeq
the analogous expression for the normal bundle. Thus the anomalous D-brane couplings
involving a gravitational eightform are indeed seen in an explicit string scattering
computation. 

However, unlike the other terms in the scattering amplitude, \eq{trR4} contains terms
which are not accounted for by the anomalous gravitational couplings in \eq{WZaction},
such as
\beq
\frac{8\pi^8}{15} (\z_1^{\perp} \cdot k_2^{\perp})(\z_2^{\Vert}\cdot k_3^{\Vert})
                   (\z_3^{\Vert}\cdot k_4^{\Vert})(\z_4^{\Vert}\cdot k_1^{\Vert})~.
\eeq
The other terms can be obtained from this one by obvious permutations, and by taking three
`transversal' factors rather than one (the latter terms have an extra minus sign).

For these `extra' terms not to interfere with anomaly cancellation arguments, they 
(or rather, their Chern-Simons forms \cite{GHM}) had better
be invariant under coordinate transformations leaving the brane invariant, {\it i.e.}
coordinate transformations that do not mix directions tangent and perpendicular to the brane
worldvolume. Indeed, in terms of linearized curvature twoforms $R$, these terms can be
written schematically as
\beq
\tr(P_{\Vert}\,R\, P_{\perp}\,R\, P_{\perp}\,R\, P_{\perp}\,R)~,
\eeq
where $P_{\Vert}$ and $P_{\perp}$ are constant matrices projecting on indices along and
perpendicular to the brane, respectively. This is the derivative of 
\beq \label{CS}
\tr(P_{\Vert}\,\omega\, P_{\perp}\,R \,P_{\perp}\,R\, P_{\perp}\,R)~,
\eeq
where $\omega$ is the (linearized) spin connection. Under coordinate transformations that do
not mix directions tangent and perpendicular to the brane, the variation of $\omega$ is a
block-diagonal matrix, such that the variation of \eq{CS} vanishes indeed.  

For orientifolds, the above analysis goes through practically unaltered. The only
difference, except for the explicit factor $-2^{p-4}$ in \eq{Oaction}, is a change in
the integrals providing the factors $\frac{14\pi^8}{45}$ and $\frac{2\pi^8}{45}$ in
\eq{trR4}. These numbers are multiplied by $-\frac{7}{8}$ 
\cite{stefanski}, such that the eightform gravitational
couplings differ only by a global factor $7.2^{p-7}$ from the D-brane ones.

\medskip
\section*{Acknowledgments}
We would like to thank Marco Bill{\'{o}} and Walter Troost for interesting discussions.
This work was supported by the European Commission TMR programme ERBFMRX-CT96-0045.

\end{document}